\documentclass[reprint,aps,prd,floats,floatfix,amsmath,amssymb,amsfonts,nofootinbib,longbibliography,superscriptaddress, twocolumn]{revtex4}

\usepackage{graphicx}
\usepackage{dcolumn}
\usepackage{bm}
\usepackage{natbib}
\usepackage{float}
\usepackage[hyperindex,colorlinks]{hyperref}
\usepackage{cleveref}
\usepackage{orcidlink}
\DeclareFontFamily{U}{mathb}{\hyphenchar\font45}
\DeclareFontShape{U}{mathb}{m}{n}{
      <5> <6> <7> <8> <9> <10> gen * mathb
      <10.95> mathb10 <12> <14.4> <17.28> <20.74> <24.88> mathb12
      }{}

\allowdisplaybreaks

\newcommand{\Khat}{\widehat K}
\newcommand{\Kdisp}{\widehat K_{\mathrm{disp}}}

\begin{document}

\title{Metric and Dispersive Dynamics of Phase Perturbations in Bohmian Quantum Mechanics}

\author{Felipe S. Escórcio\,\orcidlink{0009-0007-6723-2657}}
\email{felipe.escorcio@edu.ufes.br}
\affiliation{%
PPGFis, Universidade Federal do Espírito Santo, 29075-910, Vitória, ES, Brazil.}

\author{Hermano Velten\,\orcidlink{0000-0002-5155-7998}}%
\email{hermano.velten@ufop.edu.br}
\affiliation{%
Departamento de F\'isica, Universidade Federal de Ouro Preto, Campus Universit\'ario Morro do Cruzeiro, 35.402-136, Ouro Preto, Brazil}%
\affiliation{%
Instituto de Astronomía Teórica y Experimental, CONICET-UNC, Laprida 854, X5000BGR, Córdoba, Argentina}%

\date{\today}

\begin{abstract}
By linearizing the hydrodynamic equations associated with the Madelung transformation in the context of Bohmian dynamics, we derive an effective metric for the phase perturbation around a background state. We show that the metric previously obtained in \cite{s2026metric} by the same authors arises naturally as a particular case of a more general structure, in which the complete functional variation of the quantum potential is retained and external classical potentials can be incorporated. This generalization preserves the dependence of the quantum potential on the density and its spatial derivatives, generating additional terms that introduce dispersive corrections and act as sources in the effective Klein-Gordon equation. To organize these contributions, we introduce differential operators that separate the local sector responsible for the effective geometry from the corrections associated with the complete functional structure of the quantum potential. We identify the conditions under which these additional terms become negligible, recovering homogeneous propagation and the effective geometry for the Gaussian wave packet of a free particle.
\end{abstract}

\maketitle

\section{Introduction}
The identification of geometric structures in quantum mechanics appears in different contexts and with different meanings. A classical example is the geometry of state space, in which the structure of Hilbert space naturally induces a Riemannian metric on projective space \cite{provost1980riemannian}. In another direction, there are formulations in which aspects of quantum dynamics itself are incorporated into an effective geometric structure, either through a geometric reinterpretation of the quantum potential \cite{tavernelli2016geometrization, mikovic2025generalization, jalalzadeh2019bohmian, shojai1998geometrization}, or through generalizations based on Finsler geometry applied to the Madelung hydrodynamic representation \cite{liang2019finslerian}.

A distinct perspective arises when considering the dynamics of small perturbations around a background solution. In a broad class of systems, the linearization of the equations of motion leads to a differential operator whose principal component can take the form of a Klein--Gordon equation in an effective curved spacetime. In such cases, the coefficients of such operator (at first order in the perturbative expansion) can be interpreted as defining an effective metric. Of course  under appropriate conditions of nondegeneracy and signature. The associated causal structure determines the characteristic cones of the perturbations and may, in certain configurations, give rise to surfaces analogous to horizons. This is the central idea of the analogue gravity program \cite{barcelo2001analogue,barcelo2005analogue}. 

In de Broglie--Bohm mechanics, this approach is particularly natural because the phase of the wave function directly determines the velocity field, while the amplitude enters the dynamics through the quantum potential \cite{bohm1952suggested}. Thus, a perturbation of the phase modifies the velocity field and, consequently, the Bohmian trajectories, while perturbations of the amplitude modify the quantum potential and backreact on the evolution of the phase. The linearization of this coupled system around a background solution therefore provides a natural starting point for investigating whether the propagation of perturbations admits a description in terms of an effective geometry and, if so, which causal structure is selected by the Bohmian background. Even the de Broglie-Bohm standing as an alternative interpretation to quantum mechanics, the Bohmian framework have been widely explored in the recent literature 
\cite{durr2009bohmian,QuantumWorkPRA,capistrano,Oianguren-Asua:2022sjl,Darrow:2024zuv,Drezet:2024hqs,Drezet:2025vcv,Hatifi:2024pks,Pinto-Neto:2013toa,Pinto-Neto:2021jko}. This motivates us to explore whether Bohmian perturbations admit a similar effective geometric description as described previously.

It is in this sense that an effective metric for phase perturbations of a free particle was recently obtained in \cite{s2026metric} by the same authors. The extension of this interpretation to a formal characterization of the decoupling and eventual coupling between wavefunction branches, however, still requires further analysis. For a Gaussian wave packet of a particle, the formalism allows to study null curves and a surface with horizon-like behavior in the sector considered. The construction assumes, however, a restricted family of densities for which the variation of the quantum potential can be represented locally by the variation of the density. Outside this sector, the dependence of the quantum potential on the spatial derivatives of the density produces additional terms in the linearized equation. It therefore becomes necessary to determine how these terms modify the propagation and under which conditions the previous metric description remains valid. 

Understanding that the findings of \cite{bohm1952suggested} represent an interesting approach to the problem, but still incomplete, in this work, we generalize this construction by retaining the complete functional variation of the quantum potential and allowing for an external classical potential. The linearization of the hydrodynamic equations separates a local contribution, which defines the effective metric under specified conditions, from the corrections associated with the spatial dependence of the quantum potential. The resulting equation for the phase perturbation thus takes the form of a Klein--Gordon equation with a source.

Initially, we present the hydrodynamic formulation and establish the linearization around a general background state, explicitly retaining the functional dependence of the quantum potential on the density and its spatial derivatives. We then isolate the local sector of the perturbed equation and show under which conditions it can be identified with a Klein--Gordon operator associated with an effective metric. The remaining contributions are then organized as source terms, allowing us to distinguish the corrections arising from the complete functional structure of the quantum potential from those associated with a possible perturbation of the external classical potential. We also discuss separately the case in which the external potential is kept fixed, in which its presence modifies the background state and the coefficients of the effective geometry, without directly introducing an additional external source. Finally, we apply the formalism to the free Gaussian wave packet and identify the perturbation sector in which the corrective terms vanish, recovering the homogeneous equation and the metric construction obtained previously in Ref. \cite{s2026metric}.

\section{Linearized quantum-potential operator and dispersive correction}
\label{sec:full_operator}
We take $S$ with dimensions of action and define the velocity field $\mathbf v_0=\nabla S_0/m$, the total derivative $D_0\equiv \partial_t+\mathbf v_0\cdot\nabla$ and the background (the zeroth order in expansion series) density $\rho_0>0$. Considering the quantum potential $Q$ in its usual form
\begin{equation}
    Q\equiv Q[\rho]=-\frac{\hbar^2}{2m}
    \frac{\nabla^2\sqrt{\rho}}{\sqrt{\rho}},
    \label{eq:quantum_potential}
\end{equation}
its linear variation around $\rho_0$ can be represented in the form
\begin{equation}
    \delta Q
    =
    -\frac{\hbar^2}{4m}
    \left[
        \frac{1}{\sqrt{\rho_0}}
        \nabla^2\left(\frac{\rho_1}{\sqrt{\rho_0}}\right)
        -
        \frac{\nabla^2\sqrt{\rho_0}}{\rho_0^{3/2}}\rho_1
    \right]
    \equiv \Khat \rho_1 .
    \label{eq:deltaQ_Khat}
\end{equation}
Expanding the derivatives, the full operator $\Khat$ can be written as a sum of two different contributions $\Khat=K_0+\Kdisp$ defined according to the relations
\begin{equation}
    K_0
    =
    \frac{\hbar^2}{4m\rho_0}\nabla^2\ln\rho_0,
    \label{eq:K0}
\end{equation}
and
\begin{equation}
    \Kdisp\rho_1
    =
    -\frac{\hbar^2}{4m}
    \nabla\cdot
    \left(
        \frac{1}{\rho_0}\nabla\rho_1
    \right).
    \label{eq:Kdisp}
\end{equation}
Thus, $K_0$ corresponds to a local contribution,
whereas $\Kdisp$ contains the spatial derivatives responsible for
dispersive propagation.

To establish a connection with the construction introduced in
Ref.~\cite{s2026metric}, we consider a family $\mathcal F$ of profiles
for which the restriction of the quantum potential defines a relation
$Q_{\mathcal F}\equiv Q_{\mathcal F}(\rho)$. We introduce therefore
\begin{equation}
    K_{\mathcal F}
    \equiv
    \left(
        \frac{\partial Q_{\mathcal F}}{\partial\rho}
    \right)_{\mathcal F},
    \qquad
    c_q^2\equiv\frac{\rho_0K_{\mathcal F}}{m}>0.
    \label{eq:KF_cq}
\end{equation}
Obviously, $K_{\mathcal F}>0$. In general, $K_{\mathcal F}$ does not necessarily coincide with the term $K_0$ appearing in
Eq.~\eqref{eq:K0}. We therefore define
\begin{equation}
    \Khat
    =
    K_{\mathcal F}+\mathcal R
    =
    K_0+\Kdisp,
    \,\,\Rightarrow\,\,
    \mathcal R\equiv\Khat-K_{\mathcal F},
    \label{eq:R_definition}
\end{equation}
where the operator $\mathcal R$ will be referred to as the dispersive correction relative
to the effective nondispersive propagation.

\section{Lineared hydrodynamical equations with an external classical potential}
\label{sec:external_potential}

In this section we adopt the standard method of adopting the conservative hydrodynamical equations as our starting point. Given the following continuity and Euler equations, respectively, let $(\rho_0,S_0)$ be the hydrodynamical background solutions for the density and the phase associated with an external potential with background solution $V_0\equiv V_0(\mathbf r,t)$. Namely, 
\begin{align}
    \partial_t\rho_0+\nabla\cdot(\rho_0\mathbf v_0)&=0,
\label{eq:background_continuity}
    \\
    \partial_tS_0+\frac{|\nabla S_0|^2}{2m}
    +Q[\rho_0]+V_0&=0.
    \label{eq:background_HJ}
\end{align}
We distinguish a fixed background potential from an independent
perturbation of the external potential by writing the following expansion series
\begin{equation}
\begin{aligned}
\rho &= \rho_0+\epsilon\rho_1+O(\epsilon^2),\\
S &= S_0+\epsilon S_1+O(\epsilon^2),\\
V_{\rm ext} &= V_0+\epsilon V_1+O(\epsilon^2).
\end{aligned}
\label{eq:external_perturbations}
\end{equation}
Using the difinition for $\hat{K}$ presented in Eq.~\eqref{eq:deltaQ_Khat}, the first-order equations (i.e., linear in $\epsilon$) are
\begin{align}
    \partial_t\rho_1
    +\nabla\cdot(\mathbf v_0\rho_1)
    +\nabla\cdot
    \left(
        \frac{\rho_0}{m}\nabla S_1
    \right)
    &=0,
    \label{eq:linear_continuity_external}
    \\
    D_0S_1+\Khat\rho_1+V_1&=0.
    \label{eq:linear_HJ_external}
\end{align}
Equivalently, with $\mathbf v_1=\nabla S_1/m$,
\begin{equation}
    \partial_t\mathbf v_1
    +(\mathbf v_0\cdot\nabla)\mathbf v_1
    +(\mathbf v_1\cdot\nabla)\mathbf v_0
    =
    -\frac{1}{m}
    \nabla\left(\Khat\rho_1+V_1\right).
    \label{eq:linear_euler_external}
\end{equation}

At this point, it is worth noting the above equations are not new in the literature, apart from the fact they have been written in terms of the $\hat{K}$ operator. 

\subsection{Propagation Under an External Potential}
\label{sec:effective_metric}

We start considering an unperturbed external potential for which $V_1=0$, the form of the propagation equation eliminates only the last term in \eqref{eq:linear_HJ_external} and \eqref{eq:linear_euler_external}. This does not mean that $V_0$ is irrelevant to propagation.
Although this term will not appear explicitly, it determines the background solution and,
therefore, modifies the background solutions for $\rho_0$, $S_0$, $\mathbf v_0$, $D_0$. Consequently,  the
operator $\Khat$ itself. Thus, the fixed external potential affects the coefficients
of the propagation equation without acting as an independent source at linear order. In particular, a relation like
\begin{equation}
    V_0(\mathbf r,t)
    =
    \mathcal V\!\left(\rho_0(\mathbf r,t)\right)
    \label{eq:V0_rho0_relation}
\end{equation}
obtained by eliminating the coordinates along a specific background
solution does not imply
$\delta V_{\rm ext}=\mathcal V'(\rho_0)\rho_1$.
For an external field whose nature is unspecified, its Eulerian perturbation may remain
$V_1=0$. A constant $V_0$ is the most trivial example of this case.

In this case, whenever $\Khat^{-1}$ is well defined on the space of perturbations
under consideration, Eq.~\eqref{eq:linear_HJ_external} yields
\begin{equation}
    \rho_1=-\Khat^{-1}D_0S_1.
    \label{eq:rho1_inverse}
\end{equation}
Substituting this result into Eq.~\eqref{eq:linear_continuity_external},
we obtain the equation
\begin{equation}
    -\partial_t\left(\Khat^{-1}D_0S_1\right)
    -\nabla\cdot
    \left(
        \mathbf v_0\Khat^{-1}D_0S_1
    \right)
    +\nabla\cdot
    \left(
        \frac{\rho_0}{m}\nabla S_1
    \right)
    =0.
    \label{eq:exact_S1_equation}
\end{equation}

Using Eq.~\eqref{eq:R_definition}, the identity 
\begin{equation}
    \Khat^{-1}
    =
    K_{\mathcal F}^{-1}
    -
    K_{\mathcal F}^{-1}
    \mathcal R\Khat^{-1}
    \label{eq:resolvent_identity}
\end{equation}
allows Eq.~\eqref{eq:exact_S1_equation} to be separated into a geometric part
and a dispersive correction. Hence we define
\begin{equation}
    \mathcal L_g S_1
    \equiv
    -\partial_t
    \left(
        K_{\mathcal F}^{-1}D_0S_1
    \right)
    -\nabla\cdot
    \left(
        \mathbf v_0K_{\mathcal F}^{-1}D_0S_1
    \right)
    +\nabla\cdot
    \left(
        \frac{\rho_0}{m}\nabla S_1
    \right).
    \label{eq:Lg_definition}
\end{equation}
Without commuting $K_{\mathcal F}^{-1}$ with the derivatives,
Eq.~\eqref{eq:Lg_definition} can be written as $\mathcal L_gS_1
    =
    \partial_\mu
    \left(
        f^{\mu\nu}\partial_\nu S_1
    \right)$, with $x^0=t$ and
\begin{equation}
    f^{\mu\nu}
    =
    \begin{pmatrix}
        -K_{\mathcal F}^{-1}
        &
        -K_{\mathcal F}^{-1}v_0^j
        \\[2mm]
        -K_{\mathcal F}^{-1}v_0^i
        &
        \displaystyle
        \frac{\rho_0}{m}\delta^{ij}
        -K_{\mathcal F}^{-1}v_0^iv_0^j
    \end{pmatrix}.
    \label{eq:fmunu}
\end{equation}

The identification
$f^{\mu\nu}=\sqrt{-g}\,g^{\mu\nu}$ provide the effective metric written in such a familiar format like
\begin{equation}
    g_{\mu\nu}
    =
    \frac{\rho_0}{mc_q}
    \begin{pmatrix}
        -\left(c_q^2- v_0^2\right)
        &
        -v_{0j}
        \\[1mm]
        -v_{0i}
        &
        \delta_{ij}
    \end{pmatrix},
    \qquad
    c_q^2=\frac{\rho_0K_{\mathcal F}}{m}.
    \label{eq:effective_metric}
\end{equation}

Substituting Eq.~\eqref{eq:resolvent_identity} into
Eq.~\eqref{eq:exact_S1_equation}, we obtain
\begin{align}
    0
    =
    \mathcal L_gS_1
    &+
    \partial_t
    \left(
        K_{\mathcal F}^{-1}
        \mathcal R\Khat^{-1}D_0S_1
    \right)
    \nonumber\\
    &+
    \nabla\cdot
    \left(
        \mathbf v_0K_{\mathcal F}^{-1}
        \mathcal R\Khat^{-1}D_0S_1
    \right).
    \label{eq:Lg_plus_dispersion}
\end{align}
It is also important to realize that since the following relations hold
\begin{equation}
    \mathcal L_gS_1
    =
    \sqrt{-g}\,\Box_gS_1,
    \qquad
    \Box_g
    \equiv
    \frac{1}{\sqrt{-g}}
    \partial_\mu
    \left(
        \sqrt{-g}\,g^{\mu\nu}\partial_\nu
    \right),
    \label{eq:boxg_definition}
\end{equation}
we can define
\begin{align}
    \mathcal D S_1
    \equiv
    \frac{1}{\sqrt{-g}}
    \bigg[
    &\partial_t
    \left(
        K_{\mathcal F}^{-1}
        \mathcal R\Khat^{-1}D_0S_1
    \right)
    \nonumber\\
    &+
    \nabla\cdot
    \left(
        \mathbf v_0K_{\mathcal F}^{-1}
        \mathcal R\Khat^{-1}D_0S_1
    \right)
    \bigg].
    \label{eq:D_definition}
\end{align}
The propagation equation therefore takes the form
\begin{equation}
        \Box_gS_1+\mathcal DS_1=0.
    \label{eq:effective_wave_equation}
\end{equation}
Eq.~\eqref{eq:effective_wave_equation} is exact on the domain where the
inverse operators appearing in Eqs.~\eqref{eq:rho1_inverse} and
\eqref{eq:resolvent_identity} are well defined. The effective metric
\eqref{eq:effective_metric} describes the sector associated with
$K_{\mathcal F}$, while $\mathcal D$ contains the residual contribution of the
full quantum potential operator.

\subsection{Response to an External-Potential Perturbation}
\label{subsec:density_elimination_external}

On the domain where $\Khat^{-1}$ is well defined,
Eq.~\eqref{eq:linear_HJ_external} gives
\begin{equation}
    \rho_1
    =
    -\Khat^{-1}
    \left(
        D_0S_1+V_1
    \right).
    \label{eq:rho1_external}
\end{equation}
Consequently, substituting Eq.~\eqref{eq:rho1_external} into
Eq.~\eqref{eq:linear_continuity_external}, we obtain
\begin{align}
    &-\partial_t
    \left[
        \Khat^{-1}
        \left(D_0S_1+V_1\right)
    \right]
    \nonumber\\
    &\quad
    -\nabla\cdot
    \left[
        \mathbf v_0
        \Khat^{-1}
        \left(D_0S_1+V_1\right)
    \right]
    +
    \nabla\cdot
    \left(
        \frac{\rho_0}{m}\nabla S_1
    \right)
    =0.
    \label{eq:S1_external_full}
\end{align}
Separating the terms that depend on $S_1$ from those produced by the an
external perturbation, we obtain
\begin{align}
    &-\partial_t
    \left(
        \Khat^{-1}D_0S_1
    \right)
    -\nabla\cdot
    \left(
        \mathbf v_0\Khat^{-1}D_0S_1
    \right)
    +
    \nabla\cdot
    \left(
        \frac{\rho_0}{m}\nabla S_1
    \right)
    \nonumber\\
    &\qquad =
    \partial_t
    \left(
        \Khat^{-1}V_1
    \right)
    +
    \nabla\cdot
    \left(
        \mathbf v_0\Khat^{-1}V_1
    \right).
    \label{eq:S1_external_source}
\end{align}
The left-hand side of Eq.~\eqref{eq:S1_external_source} is the
same operator that appears in the homogeneous equation
\eqref{eq:exact_S1_equation}, while the right-hand side
constitutes an explicit new source for $S_1$ due to the perturbed external potential contributions. 

Using the identity
\eqref{eq:resolvent_identity},
\begin{equation}
    \Khat^{-1}V_1
    =
    K_{\mathcal F}^{-1}V_1
    -
    K_{\mathcal F}^{-1}
    \mathcal R\Khat^{-1}V_1,
    \label{eq:external_source_decomposition}
\end{equation}
we are able to write down a more general relation in such way that Eq.~\eqref{eq:S1_external_source} gives rise to
\begin{align}
    &\mathcal L_gS_1
    +
    \partial_t
    \left(
        K_{\mathcal F}^{-1}
        \mathcal R\Khat^{-1}D_0S_1
    \right)
    \nonumber
    +
    \nabla\cdot
    \left(
        \mathbf v_0
        K_{\mathcal F}^{-1}
        \mathcal R\Khat^{-1}D_0S_1
    \right)
    \nonumber\\
    &=
    \partial_t
    \left(
        K_{\mathcal F}^{-1}V_1
    \right)
    +
    \nabla\cdot
    \left(
        \mathbf v_0K_{\mathcal F}^{-1}V_1
    \right)
    -
    \partial_t
    \left(
        K_{\mathcal F}^{-1}
        \mathcal R\Khat^{-1}V_1
    \right)
    \nonumber\\
    &\quad
    -
    \nabla\cdot
    \left(
        \mathbf v_0
        K_{\mathcal F}^{-1}
        \mathcal R\Khat^{-1}V_1
    \right).
    \label{eq:S1_external_geometric}
\end{align}
The right-hand side corresponds to the response associated with the
local sector $K_{\mathcal F}$ and includes, through the last two contributions,
the residual response governed by $\mathcal R$. In particular,
even if $V_1(\mathbf r,t)$ is a local function,
$\Khat^{-1}V_1$ and
$\mathcal R\Khat^{-1}V_1$ may be nonlocal. This nonlocality
arises from inverting the differential operator $\Khat$ and not from a
modification of the quantum potential itself.

Dividing Eq.~\eqref{eq:S1_external_geometric} by $\sqrt{-g}$ and using
Eqs.~\eqref{eq:boxg_definition} and \eqref{eq:D_definition}, we obtain
\begin{equation}
    \Box_gS_1+\mathcal DS_1
    =
    \mathcal J_V,
    \label{eq:wave_equation_external_source}
\end{equation}
where the external source is
\begin{align}
    \mathcal J_V
    \equiv
    \frac{1}{\sqrt{-g}}
    \bigg\{
    \partial_t
    \left(
        K_{\mathcal F}^{-1}V_1
    \right)
    +
    \nabla\cdot
    \left(
        \mathbf v_0K_{\mathcal F}^{-1}V_1
    \right)
    \nonumber\\
    -
    \partial_t
    \left(
        K_{\mathcal F}^{-1}
        \mathcal R\Khat^{-1}V_1
    \right)-
    \nabla\cdot
    \left(
        \mathbf v_0
        K_{\mathcal F}^{-1}
        \mathcal R\Khat^{-1}V_1
    \right)
    \bigg\}.
    \label{eq:JV_definition}
\end{align}
Therefore, an independent
perturbation $V_1$ appears as a source in the effective equation \eqref{eq:wave_equation_external_source}.

It is noted that the perturbation of the external potential induces a phase response through the source term and the initial data compatible with the linearized equations. By the Bohmian guidance law, the gradient of this response modifies the velocity field. The change in the trajectories is then determined by integrating this field along the background flow, including the effect of spatial variations in the background velocity on the displacement. Thus, the equation for \(S_1\), together with the guidance law, establishes an explicit relation between the external perturbation and the response of the trajectories.


\subsection{A particular case: Recovery of the Gaussian sector for a free particle}
\label{subsec:gaussian_sector}

Consider the Gaussian family of solutions in which $\rho_0(\mathbf r)
    =
    A e^{-\beta r^2}$ with $A= \left(\beta/\pi
    \right)^{N/2}$,
and an infinitesimal translation with fixed $\beta$,
\begin{equation}
    \rho_\epsilon(\mathbf r)
    =
    \rho_0(\mathbf r-\epsilon\mathbf u),
    \qquad
    Q_\epsilon(\mathbf r)
    =
    Q_0(\mathbf r-\epsilon\mathbf u).
    \label{eq:gaussian_translation}
\end{equation}
To first order,
\begin{equation}
    \rho_1
    =
    2\beta(\mathbf u\cdot\mathbf r)\rho_0.
    \label{eq:gaussian_rho1}
\end{equation}
For the proposed density, the background contribution of the quantum potential can be determined as
\begin{equation}
    Q_0
    =
    \frac{\hbar^2}{2m}
    \left(
        N\beta-\beta^2r^2
    \right),
\end{equation}
and therefore,
\begin{equation}
    \delta Q
    =
    \frac{\hbar^2\beta^2}{m}
    (\mathbf u\cdot\mathbf r).
    \label{eq:gaussian_deltaQ}
\end{equation}
On the other hand, along this same family,
\begin{equation}
    K_{\mathcal F}
    =
    \frac{\hbar^2\beta}{2m\rho_0},
\end{equation}
so that Eqs.~\eqref{eq:gaussian_rho1} and
\eqref{eq:gaussian_deltaQ} imply
\begin{equation}
    \Khat\rho_1
    =
    \delta Q
    =
    K_{\mathcal F}\rho_1.
    \label{eq:gaussian_Khat}
\end{equation}
Using Eq.~\eqref{eq:R_definition}, it follows immediately that $\mathcal R\rho_1=0$. Thus, for Gaussian translational perturbations that satisfy the
linearized equations,
\begin{equation}
    \mathcal R\Khat^{-1}D_0S_1=0,
\end{equation}
causing the dispersive correction \eqref{eq:D_definition} to vanish, $\mathcal DS_1=0$.
The effective equation therefore reduces to the expected result
\begin{equation}
    \Box_gS_1=0.
    \label{eq:gaussian_wave_equation}
\end{equation}
Since the correction $\mathcal R$ vanishes in this sector, the linearized dynamics is described
entirely by the effective geometry associated with the metric
\eqref{eq:effective_metric}, leading naturally to the particular results described in Ref. \cite{s2026metric}.

\section{Conclusions}
We have generalized the main results of Ref. \cite{s2026metric} about the linearization procedure of the hydrodynamic equations in the context of de Broglie--Bohm mechanics. Specifically, our calculations in the present work preserve the full functional dependence of the quantum potential and additionally allows for the presence of an external classical potential. We show that the dynamics of phase perturbations can be decomposed into a local sector, described by an effective metric, and additional contributions arising from the spatial dependence of the quantum potential. These contributions appear as source terms in the effective Klein--Gordon equation and may generate both dispersive corrections and a nonlocal dependence of the perturbation itself. When the classical potential is also perturbed, an additional external source term arises; when it is kept fixed, its influence remains incorporated into the background state and, consequently, into the coefficients that define the effective metric.

The effective geometry should therefore be interpreted as a structure governing the propagation of phase perturbations, rather than as a direct geometrization of Bohmian trajectories. Nevertheless, the formulation developed here provides a natural framework for investigating under what conditions, in the presence of the full form of the quantum potential, these perturbations can reach the effectively occupied configuration and produce a local modification of the phase field entering the guidance equation. A formal demonstration of this possibility, however, remains open and constitutes a relevant direction for future work. This issue requires, in particular, carefully distinguishing the curves associated with the propagation of perturbations from the Bohmian trajectories themselves: even when a perturbation reaches the region of the occupied configuration, an effective deviation of the Bohmian dynamics will occur only if it produces a non-negligible contribution to the phase gradient evaluated at that configuration.

Finally, we identify the sector corresponding to the free Gaussian wave packet in which the corrective terms become negligible, so that the homogeneous equation and the effective geometry previously obtained in \cite{s2026metric} are recovered. This result explicitly establishes the regime in which the previous construction emerges from the more general formulation presented here and, at the same time, situates that effective geometry as a particular case of a broader perturbative structure.

\acknowledgments{The authors thank the Brazilian funding agencies FAPES, CAPES, and CNPq (project 201644/2026-6) for partial financial support. Discussions with Julio César Fabris on this topic are highly appreciated.}

\onecolumngrid

%








\end{document}